\documentclass[aip,jcp,numerical,reprint]{revtex4-1}
\usepackage{graphicx}
\usepackage{amsfonts}
\usepackage{amsmath}
\usepackage[english]{babel}
\usepackage{multirow}
\def\Pe{\ensuremath{\text{\textit{Pe}}}}
\def\kT{\ensuremath{k_\text{B}T}}

\usepackage[usenames,dvipsnames]{xcolor}
\usepackage[normalem]{ulem}

\begin{document}
\title{On the Bauschinger effect in supercoooled melts under shear:
results from mode coupling theory and molecular dynamics simulations}
\date{\today}
\def\unikn{\affiliation{Fachbereich Physik,
  Universit\"at Konstanz, 78457 Konstanz, Germany}}
\def\dlr{\affiliation{Institut f\"ur Materialphysik im Weltraum,
  Deutsches Zentrum f\"ur Luft- und Raumfahrt (DLR), 51170 K\"oln, Germany}}
\def\zk{\affiliation{Zukunftskolleg,
  Universit\"at Konstanz, 78457 Konstanz, Germany}}
\def\unidus{\affiliation{Institut f\"ur Theoretische Physik II,
  Soft Matter, Heinrich-Heine-Universit\"at D\"usseldorf,
  40225 D\"usseldorf, Germany}}
\author{Fabian Frahsa}\unikn
\author{Amit Kumar Bhattacharjee}
\thanks{Present address: Fachbereich Physik,
  Universit\"at Konstanz, 78457 Konstanz, Germany.}\dlr
\author{J\"urgen Horbach}
\thanks{Present address: Institut f\"ur Theoretische Physik II,
  Soft Matter, Heinrich-Heine-Universit\"at D\"usseldorf,
  40225 D\"usseldorf, Germany.} \dlr
\author{Matthias Fuchs} \email{Matthias.Fuchs@uni-konstanz.de} \unikn
\author{Thomas Voigtmann}\unikn\dlr\zk

\begin{abstract}
We study the  nonlinear rheology of a glass-forming binary mixture under the reversal of shear 
flow using molecular dynamics simulations and a schematic model of the mode-coupling theory
of the glass transition (MCT). Memory effects lead to a history-dependent response, as
exemplified by the vanishing of a stress-overshoot phenomenon in the stress--strain curves
of the sheared liquid, and a change in the apparent elastic coefficients around states
with zero stress. We investigate the various retarded contributions to the stress response
at a given time schematically within MCT. The connection of this macroscopic response to
single-particle motion is demonstrated using molecular-dynamics simulation.
\end{abstract}

\maketitle

\section{Introduction}

Dense liquids exhibit slow relaxation processes in thermal equilibrium,
characterized by a time scale $\tau$ that diverges as the liquid is
driven to structural arrest into an amorphous solid \cite{KobReview}.
This opens a window rich in nonlinear phenomena under flow, both amenable to
experiment and theoretical treatment. It is of interest, because the externally imposed flow strongly disturbs
the equilibrium relaxation processes. More precisely, given a typical
single-particle relaxation time scale $\tau_0$, we consider shear flow
with deformation rate $\dot\gamma$ such that $\Pe_0=\dot\gamma\tau_0\ll1$,
but $\Pe=\dot\gamma\tau\gg1$. This nonlinear rheology has in particular
been studied in colloidal dispersions and soft-matter systems, where
the mesoscopic size of the constituent particles implies relaxation times
of $\mathcal O(\text{s})$, easily probed by moderate flow in experiment.
In the flowing steady state,
shear thinning is the most prominent phenomenon: \cite{larson,oswald} the apparent
viscosity of the sheared fluid decreases rapidly with increasing
shear rate $\dot\gamma$.
Quite generally, the flowing steady state is characterized by a nonlinear relation between the
macroscopic stress $\sigma$ and 
the applied shear rate $\dot\gamma$.

Going beyond the steady state, the behavior under start-up flow has received
particular attention. As a steady shear flow of rate $\dot\gamma$ is
suddenly switched on at $t=0$, stresses $\sigma$
increase from zero (assuming the experiment was started in an
equilibrated and hence stress-free configuration) to their steady-state
value $\sigma(\dot\gamma)$. The resulting stress-strain curve $\sigma(\gamma)$,
measured as a function of applied deformation $\gamma=\dot\gamma t$ for
$t>0$,
exhibits a peculiar intermediate maximum, termed stress overshoot.
This is typically found in amorphous metals \cite{Schuh.2007},
or (for different reasons) polymer melts \cite{DealyLarson}.
The maximum of the $\sigma(\gamma)$ curve has been linked to
material stability: it can be taken as a yield stress that separates
reversible (linear and nonlinear) elastic deformation at small strains from irreversible plastic
flow induced by larger deformations.
More recently, stress overshoots have also been discussed in
soft colloidal systems, underlining a generic mechanism applicable
in a wide variety of system classes \cite{Zausch2008,Koumakis2012,Siebenbuerger2012,Amann2012}.

As a transient phenomenon, the stress overshoot is susceptible to
the history of sample preparation. Molecular dynamics simulations have
addressed this focusing on the shear-molten amorphous solid, where
a systematic dependence was found on the waiting time $t_{\rm age}$ between sample
preparation and flow start-up \cite{vaboba}, as long as the start-up configuration is
not fully equilibrated, $t_{\rm age}\ll\tau$.

In the engineering literature it is well established that the stress-strain
curve depends on the deformation history of the sample. One particular example
was found empirically by Johann Bauschinger in the 1880's during experiments
on various steels \cite{Bauschinger}: these materials typically are softer under
compressional load after having been subject to tension. This is, in the strict
sense, called the Bauschinger effect. As an abstraction, take the case
where pre-shear is applied in one particular direction, and consider the
modified stress--strain curve measured if shear is then applied in the opposite
direction. Typically, the regime of elastic deformation shrinks, so that plastic
rearrangements dominate much earlier than in an initially unstrained sample.

Reversing the shear rate, which we will call 'Bauschinger effect' also, in the following, has been discussed \cite{Langer2008} in the framework of the shear-transformation-zone theory (STZ),
where one attributes internal degrees of freedom to the STZs, which carry the memory of the deformation history.
Recent simulation studies of the Bauschinger effect \cite{Karmakar2010}
in a system quenched into its low-temperature glassy state have been
analyzed in terms of anisotropic elastic constants that arise due to
the deformation history.

It is a matter of current debate whether and how the rheological properties of
amorphous materials deep in the glassy state (here referred to as essentially
the $T\to0$ case), and at much higher temperaturs in the melt, are related.
In athermal systems, concepts like STZs \cite{Langer2008} and avalanches of yielding
events \cite{Lemaitre} are widely recognized as useful to
rationalize macroscopic behavior. But the range of applicability of
these approaches is not well understood when dealing with
liquids close to their glass transition, where thermal fluctuations dominate
yielding mechanisms that may be qualitatively different from, say,
athermal jamming
\cite{Voigtmann.2011}. The rheology of dense liquids
is the realm of a recent extension
of the mode-coupling theory of the glass transition (MCT) to sheared colloidal
systems \cite{Fuchs2002,FuchsCates2009,Brader2012}. Intriguingly, some of the
macroscopic phenomena are quite alike in both quasi-athermal systems and
liquids driven by thermal noise.

Here we investigate, both by mode-coupling theory and computer simulation,
the nonlinear rheology under flow reversal in the
initially equilibrated liquid state, without referring to $T\to0$ properties.
The initial stress overshoot is shown to vanish if the flow is reversed
in the steady state. Hence, the Bauschinger effect does not require an
explanation associated with
properties of the quenched glass or aging phenomena.
We study its
connection to the time-dependent memory of past flow that is typical
for dense liquids, by investigating the effect of flow reversal applied
at various times during the initial evolution from equilibrium to steady
state. A clear separation between reversible elastic rearrangements and
irreversible plastic flow emerges. We connect the macroscopic stress
response to microscopic particle motion, quantified through the transient
mean-squared displacements.

The paper is organized as follows:  Section~\ref{sec:1} contains the exposition
of the theoretical approach, and section~\ref{sec:2} gives an overview
of the simulation.
In Sec.~\ref{sec:3} we discuss the comparison between theory and simulation, and the insights thereby gained. Section~\ref{sec:4} concludes with an outlook.

\section{\label{sec:1}Theory}

Our theoretical approach is based on  the mode-coupling theory of the glass transition (MCT)
\cite{Gotze2009},
as extended to deal with nonlinear rheology in the integration-through transients (ITT) formalism
\cite{Fuchs2002,Brader2012}. ITT-MCT is a microscopic (albeit approximate)
theory describing
 the dynamical correlation functions of the wave-vector dependent density fluctuations
in glass-forming liquids. Suitably generalized
Green-Kubo relations provide the nonequilibrium transport coefficients in the nonlinear-response
regime; within MCT, they are expressed as integrals over the density correlators. The coupling
coefficients of the theory are fully determined by knowing the equilibrium static structure of
the liquid.
The theory then predicts an ideal glass transition to occur as a bifurcation transition: smooth changes
in control parameters such as density or temperature bring about a discontinuous change in the
long-time behavior of the density correlation functions. These decay to zero in the equilibrium
liquid, but attain a finite long-time limit, the glass form factor, in the ideal glass. The theory compares semi-quantitatively to the linear rheology measured in colloidal hard sphere dispersions \cite{FuchsBallauff,Siebenbuerger2009,Siebenbuerger2012b}.

Since the full microscopic theory is rather cumbersome to solve, various simplifications have
been devised. Schematic models try to capture mathematically the essence of the bifurcation
transition by reducing the number of correlation functions to a few or only one, at the cost
of neglecting spatial information.  The important correlations contained in the schematic models are temporal ones.  Memory effects are contained in generalized friction kernels depending on (up to) three times. Causality is automatically enforced, and the flow history of the sample is contained in the friction-kernels via time- or strain-dependent elastic coefficients and structural relaxation functions. The latter describe plastic deformation, and their nonlinear equations of motion are the heart of the theory. The non-Markovian equations of the schematic models contain a small number of fitting parameters that mimic
the relation between the equilibrium structure (encoding the interaction potential)
and the theory's coupling coefficients. Smooth and regular variations of the parameters lead to dramatic elasto-plastic and viscoelastic changes in the structural relaxation and non-linear rheology.  The interpretation of the effects in terms of spatially dependent particle rearrangements can not be achieved by the schematic model, and thus it remains unresolved to which degree 'force-chains', 'heterogeneities', or specific 'plastic rearrangement centers' are described in a (presumably) space-averaged manner.

\subsection{Schematic MCT Model}


A schematic MCT model for the time-dependent rheology of colloidal suspensions was introduced
in Ref.~[\onlinecite{Brader2009}]; it has been used to study large-amplitude oscillatory shear
\cite{Brader2010a}, mixed flows 
\cite{Farage2012},
 and double-step strain \cite{Voigtmann2012}, and generalizes a model used to analyze stationary flow curves \cite{FuchsBallauff,Siebenbuerger2009}.
In the model, the nonlinear generalization of the Green-Kubo relation for the shear-stress tensor element
$\sigma\equiv\sigma_{xy}$ reads, imposing spatially homogeneous but otherwise arbitrary simple shear flow
with rate $\dot\gamma$ in the $x$-direction, with gradient in the $y$-direction,
\begin{equation}
  \sigma(t)= \int\limits_{-\infty}^{t} \!\text{d}t'\dot\gamma(t') G(t,t',[\dot\gamma])\,.\label{eqn:greenkubo}
\end{equation}
The generalized shear modulus $G(t,t',[\dot\gamma])$ depends on the full flow history encoded
in the time-dependent shear rate $\dot\gamma(t)$. Outside the steady state, it also depends on
the two times corresponding to the underlying fluctuations separately.
In the spirit of MCT, we approximate the shear modulus in terms of the transient density correlation
function $\Phi(t,t')$,
\begin{equation}\label{eqn:gschem}
  G(t,t')=v_{\sigma}(t,t')\, \Phi^2(t,t').
\end{equation}
Here, an elastic coefficient $v_\sigma$ appears that was set to a constant in the original formulation of
the model \cite{Brader2009}. It measures the strength of stress fluctuations caused by pair-density fluctuations. The
microscopic theory implies a relation
between $\Phi(t,t')$ and $G(t,t')$ that involves an integration over wave vectors, including
nontrivial weights that depend on time since density fluctuations are advected by shear.
It is therefore plausible to extend the original schematic model by allowing for a
prefactor $v_\sigma(t,t')$ that depends on time through the accumulated strain
$\gamma(t,t')=\int_{t'}^t\dot\gamma(s)\,\text{d}s$,
\begin{equation}
  v_{\sigma}(t,t')=v_{\sigma}^*\, \cdot\left( 1-\left(\frac{\gamma\left(t,t'\right)}{\gamma^{*}}\right)^2\right)\, \exp\left(-\left(\frac{\gamma\left(t,t'\right)}{\gamma^{**}}\right)^2\right).\label{eqn:ammanprefactor}
\end{equation} 
Symmetry dictates that the direction of the strain does not matter, leading to an even function
$v_\sigma=v_\sigma(\gamma(t,t')^2)$.
Equation \eqref{eqn:ammanprefactor} generalizes the time-dependent coupling first introduced for constant shear rates in Ref.~[\onlinecite{Amann2012}], where it was justified by a comparison with fully microscopic ITT-MCT calculations in two dimensions.
The functional form is chosen simple enough for rapid numerical implementation and allows for a quantitative discussion of stress overshoots in model colloidal suspensions
\cite{Amann2012, MarcoLauratiKJMutchNKoumakisJZauschCPAmannAndrewBSchofieldGeorgePetekidisJohnFBradyJuergenHorbach}.  
The parameter $v_{\sigma}^*$ sets the scale of the generalized shear modulus. 
We obtain it by matching it to the linear-response low-frequency plateau modulus $G_{\infty}$ of
the simulation.  The shear elastic constant $G_\infty$ itself often is called `shear modulus', and can,
e.g., be measured in the low-frequency linear elastic modulus of the glass, $G'(\omega\to0)=G_\infty$,
or as one of the two Lam\'e-coefficients of an isotropic solid \cite{Klix2012}. 
At a critical strain scale $\gamma^{\ast}$, the time-dependent modulus becomes negative and the 
stress-strain curve exhibits an overshoot with its peak at the critical strain. 
$\gamma^{\ast\ast}$ describes the decay of the overshoot. 
To reduce the number of free parameters, we keep it linearly proportional \cite{Amann2012} to $\gamma^*$.

The time-dependent elastic coefficient given in Eq.~\eqref{eqn:ammanprefactor} is carefully chosen to become
negative in a small strain window relevant for the decay of the correlation functions.
Doing so, it allows to circumvent an obvious limitation of the schematic model: setting
$v_\sigma=\text{const.}$, Eq.~\eqref{eqn:gschem} has a definite sign. In startup shear,
cf.\ Eq.~\eqref{eqn:greenkubo}, this implies that $\sigma(t)$ varies monotonically with
accumulated strain. The original schematic model hence does not describe stress overshoot
phenomena. Motivated by the full microscopic expression of ITT-MCT \cite{FuchsCates2009,Amann2012}
and the observation that the wave-vector dependent MCT, even with further isotropy assumptions,
produces a stress overshoot through a small negative dip in $G(t,t')$, cf.\ Ref.~[\onlinecite{Zausch2008}],
Eq.~\eqref{eqn:ammanprefactor} is a simple way to incorporate the phenomenology.
In the following, we take it as one established way to match the startup stress--strain curves (see below for a comparison to simulation), and discuss the consequences of the history-integral formulation of Eq.~\eqref{eqn:greenkubo} under flow reversal.

The schematic density correlator is the solution of a Mori-Zwanzig-like memory equation
\begin{equation}\label{eq:eom}
 \dot\Phi(t,t') + \Gamma\left(\Phi(t,t') + \int\limits_{t'}^{t}\!\text{d}t''\, m(t,t'',t')\dot\Phi(t'',t')\right) =0,
\end{equation}
with initial condition $\Phi(t,t')=1-\Gamma(t-t')+\ldots$.
Here $\Gamma$ denotes the initial decay rate of the density correlator, 
which corresponds to a reciprocal microscopic relaxation time. We take
$\Gamma$ to be independent of the shear rate.

To close this equation of motion, MCT approximates the memory kernel $m$ through
a nonlinear polynomial of the correlation functions. In contrast to a Markovian approximation, this postulates that the kernel relaxes on the same time scale as the correlator, and that both need to be obtained self-consistently. Conceptually, this is motivated by the assumption that stress fluctuations captured in $m$  arise from slow structural rearrangements, which themselves are captured in $\Phi$. The simplest schematic model
that recovers the asymptotic behavior found for typical quiescent glass-forming liquids
is the F$_{12}$ model \cite{Gotze2009}, whose extension to time-dependent shear
reads
\begin{equation}\label{eqn:mem}
m(t,t'',t')=h(t,t'')\, \left[v_1\Phi(t,t'')+v_2\Phi^2(t,t'')\right].
\end{equation}
The parameters $v_1$ and $v_2$ are the coupling coefficients driving the system
through the glass transition. There is a line of bifurcation points $(v_1^c,v_2^c)$, and we pick
$v_2=v_2^c$ and $v_1=v_1^c+\epsilon/(\sqrt2-1)$ with $v_1^c=2(\sqrt2-1)$ as in
previous work. The parameter $\epsilon$ controls the distance to the glass transition:
$\epsilon<0$ corresponds to liquid states, where the correlation function decays to zero
in the quiescent system, on a time scale $\tau$ that diverges as $\epsilon\to0$.
States with $\epsilon>0$ are idealized glass states, where the quiescent correlation
function attains a positive long-time limit, $\lim_{t\to\infty,t\gg t'}\Phi(t,t')=f>0$.
This value is called the non-ergodicity parameter or glass form factor. The elastic constant in the model then follows as $G_\infty=v_\sigma^* f^2$.

One of the predictions of ITT-MCT is that (steady) shear melts the glass; the correlation
functions even in the glass show shear-induced decay due to a loss of memory \cite{Fuchs2002,fuca}. In the
schematic model, this is modeled by an ad-hoc strain-reduction function,
\begin{equation}\label{eq:hfunction}
 h(t,t')= \frac{1}{1+\left(\gamma(t,t')/\gamma_c\right)^2}.
\end{equation}
Symmetry again dictates that $h$ is an even function of the accumulated strain.  
The parameter $\gamma_c$ is a critical strain and describes how fast the memory 
of the glass decays due to its deformation. In principle, the microscopic form
of the MCT equation suggests further combinations of the three times to appear
in Eq.~\eqref{eqn:mem}, as exploited for example in Refs.~[\onlinecite{Brader2009}] and [\onlinecite{Voigtmann2012}]. We
neglect these terms here, to make contact with earlier schematic-model analyses
of the nonlinear steady-state rheology of colloidal suspensions \cite{FuchsBallauff,Siebenbuerger2009,Siebenbuerger2012b}. In principle, the strain-reduction function $h$ could also contain negative regions, as arises in the stress-density coupling $v_\sigma$. This could cause the transient correlator $\Phi$ to become negative for intermediate times as seen in computer simulations  and microscopic ITT-MCT calculations in two dimensions \cite{Krueger2011}.  For simplicity, this is neglected.

\subsection{\label{subsec:3}Bauschinger effect}

\begin{figure}
\centerline{\includegraphics[width=.9\linewidth]{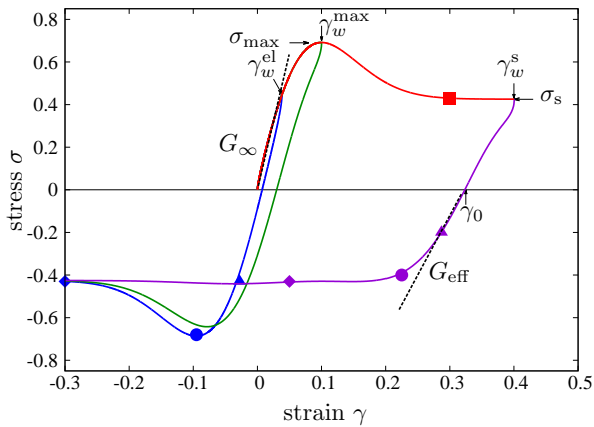}}
\caption{\label{fig:hysteresis}
  Linear plot of the
  stress-strain relation $\sigma(\gamma)$ at fixed density and
  shear rate for various flow histories: startup flow with constant shear rate (red), flow reversal in the steady state (s, purple),   from the elastic regime (el, blue), and from the point of the
  stress overshoot (max, green).  The linear increase with the quiescent elastic modulus $G_\infty$ in startup, and with $G_\text{eff}$ after shear-reversal  are indicated by dotted lines. Characteristic stress--strain points, whose transient modulus is discussed in the text, are marked by symbols: The red square gives the stationary stress. It corresponds to an integration along the line from A to B in
Fig.~\ref{fig:timeplane}. Triangles mark stresses after flow reversal in the linear regime and correspond to an integration along the line C' to E'. Discs mark stresses after flow reversal in the overshoot regime and are obtained via integrations along the
line C to E in Fig.~\ref{fig:timeplane}. }
\end{figure}

The flow history we consider in this work is given by two intervals of constant shear rate:
\begin{equation}
  \dot\gamma(t)=\begin{cases} 0 & \text{for $t< 0$,}\\
+\dot\gamma & \text{for $0<t<t_w$,}\\
  -\dot\gamma & \text{for $t>t_w$.}\end{cases}\label{shearrate}
\end{equation}
We mainly discuss a fixed value for $\dot\gamma$, and different waiting times $t_w$ over which
the ``pre-shear'' in the positive direction is applied. These times correspond to an accumulated
strain $\gamma_w=\dot\gamma t_w$. The amount of this pre-strain is decisive and determines the stress--strain curve after switching. 
Figure~\ref{fig:hysteresis} presents typical results obtained in the schematic model  as unmodified stress-versus-strain curves; it serves to introduce the flow history and characteristic parameters.  After startup of shear at $t=0$,
the system first undergoes an elastic transient, characterized by a stress--strain curve
$\sigma(\gamma)$ that is approximately linear, $\sigma\approx G_\infty\gamma$, given by the quiescent
elastic shear modulus. At the end of this regime, the $\sigma$-versus-$\gamma$ curves becomes
sublinear, until the position of the overshoot is reached at some strain $\gamma^\text{max}$,
typically of the order of $10\%$. At large strains, one reaches the steady state characterized
by a constant $\sigma(\gamma\to\infty)=\sigma_s$ independent of $\gamma$. For the flow reversal, we consider switching
the shear rate at three typical points in either regime: $\gamma_w^\text{el}$ inside the
elastic transient, $\gamma_w^\text{max}$, and $\gamma_w^\text{s}$ in the steady state. The $\gamma_w$ values are marked in Fig.~\ref{fig:hysteresis}. 
A value $\gamma_w^\text{s}=0.4$ in the schematic model turns out to be sufficient so that
no further $\gamma_w$ dependence is observed for larger values besides a trivial shift. 
To ease the interpretation of the curves at different switching times, we choose $\gamma_w^\text{el}$ such that
the corresponding stress is equal to the steady-state one,
$\sigma(\gamma_w^\text{el})=\sigma(\gamma_w^\text{s})$.
Generally, after reversing the flow direction, the stress first decreases back to zero, defining
a strain value $\gamma_0$. The stress--strain curves in later figures will be compared in a
$|\sigma|$-versus-$|\gamma-\gamma_0|$ representation.
While it does hide the ``hysteresis'' loop which can be recognized in Fig.~\ref{fig:hysteresis}, and which is familiar
in the engineering literature, it simplifies the quantitative analysis.

All calculations in the schematic model  are done with parameters that have been found typical in applications of predecessors
of the model to large amplitude oscillatory shear \cite{Brader2010a}, and especially to startup flow \cite{Amann2012,MarcoLauratiKJMutchNKoumakisJZauschCPAmannAndrewBSchofieldGeorgePetekidisJohnFBradyJuergenHorbach}. The values are:
$\Gamma=100$,  $\gamma_c=0.75$,
$\gamma^{\ast}=0.1$, and $\gamma^{\ast\ast}=0.133$; see also 
Fig.~\ref{fig:stressstrain}.
The considered state in Fig.~\ref{fig:hysteresis}, which will be analyzed in more detail in Figs.~\ref{fig:3erpanel_for_reversal_in_steadystate} and \ref{fig:3erpanel_for_reversal_in_elastic_regime}, is a glass very close to the glass transition in order to achieve
a large separation of the structural dynamics from the short-time one; $\epsilon=10^{-4}$. 
The shear rate $\dot\gamma=5\times 10^{-3}$, for which the Bauschinger effect is studied, corresponds to a rather small Peclet number, which in the model is given by Pe$_0=\dot\gamma/\Gamma$, and thus takes the value $5\times 10^{-5}$. These two values  provide access to the asymptotic regime of ITT-MCT for $\epsilon\to0$ and Pe$_0\ll1$. Slightly different $\epsilon$ and $\dot\gamma$ values are used in the comparisons to the simulations in Sec.~\ref{sec:3}.

Figure~\ref{fig:hysteresis} contains a wealth of information, which will be discussed in Sec.~\ref{sec:3} together with the simulation results.

\section{\label{sec:2}Simulation}

We perform nonequilibrium molecular dynamics computer simulation for
a glass-forming binary mixture. Particles interact through
a purely repulsive soft-sphere potential. Denoting particle species by
$\alpha,\beta$, the truncated Lennard-Jones potential due to Weeks,
Chandler, and Andersen \cite{Weeks1971} reads
\[
\label{wcapot}
  V_{\alpha\beta}^\text{WCA}(r)=\begin{cases}
    4\epsilon_{\alpha\beta}\left[\left(
    \frac{\sigma_{\alpha\beta}}{r}\right)^{12}-\left(
    \frac{\sigma_{\alpha\beta}}{r}\right)^6+\frac14\right]
    & \text{$r<r_{c,\alpha\beta}$}\\
    0 & \text{else}
  \end{cases}\,,
\]
where $r_{c,\alpha\beta}=2^{1/6}\sigma_{\alpha\beta}$ is a cutoff for
the interaction range.
To ensure continuity of force and conservation of total energy in the NVE
ensemble, we apply a smoothing function,
$V_{\alpha\beta}(r)=V_{\alpha\beta}^\text{WCA}(r)S_{\alpha\beta}(r)$ with
$S_{\alpha\beta}(r)=(r-r_{c,\alpha\beta})^4/[h^4+(r-r_{c,\alpha\beta})^4]$
with $h=10^{-2}\sigma_{\alpha\beta}$ \cite{Zausch2010}.
We choose units of energy such that $\epsilon_{\alpha\beta}=1$, and
units of length such that $\sigma_{\text{AA}}=1$. The unit of time
is given by $\sqrt{m_\text{A}\sigma^2_{\text{AA}}/\epsilon_{\text{AA}}}$
where $m_\text{A}=m_\text{B}=1$ are the masses of the particles.
For the smaller particles,
$\sigma_{\text{BB}}=5/6$, and the mixture is additive,
$\sigma_{\text{AB}}=(\sigma_{\text{AA}}+\sigma_{\text{BB}})/2$.
This system, albeit with differing masses,
has been previously studied in the quiescent state and identified
as a model glass former in its equimolar composition \cite{Hedges2007}.

The system is coupled to a dissipative particle dynamics (DPD) thermostat:
the equations of motion read \cite{Groot1997}
\begin{subequations}
\label{eom}
\begin{align}
  m_\alpha\dot{\vec r}_\alpha&=\vec p_\alpha\,,
  \\
  \dot{\vec p}_\alpha&=\vec F_\alpha^\text{C}(\{\vec r\})
  +\vec F_\alpha^\text{D}(\{\vec r,\vec p\})
  +\vec F_\alpha^\text{R}(\{\vec r\})\,.
\end{align}
\end{subequations}
where $\vec r$ and $\vec p$ are positions and momenta of the particles,
and $\vec F^i$ denotes conservative ($i=\text{C}$), dissipative
($i=\text{D}$), and random ($i=\text{R}$) forces.
With the interparticle separation $r_{ij}=|\vec r_i-\vec r_j|$ between
particles $i$ and $j$ of species $\alpha$ and $\beta$,
\begin{align}
  \vec F_\alpha^\text{C}
  &=-\sum_{j\neq i}\vec\nabla V_{\alpha\beta}(r_{ij})\\
  \vec F_\alpha^\text{D}
  &=-\sum_{j\neq i}\zeta w^2(r_{ij})(\hat{\vec r}_{ij}\cdot\vec v_{ij})
  \hat{\vec r}_{ij}\\
  \vec F_\alpha^\text{R}
  &=\sum_{j\neq i}\sqrt{2\kT\zeta}\,w(r_{ij})\mathcal N_{ij}\hat{\vec r}_{ij}
\end{align}
Here, $\vec v_{ij}=\vec p_i/m_i-\vec p_j/m_j$ is the relative velocity between
the two particles, and $\hat{\vec r}_{ij}=(\vec r_i-\vec r_j)/r_{ij}$ their
unit separation. The appearance of the relative velocity in the dissipative
force is crucial to obtain local momentum conservation and Galilean
invariance \cite{Zausch2010}. $w(r)$ is a cutoff function, set to
$w=1$ for $r<r_{c,\text{DPD}}/2$ and $w=0$ elsewhere. We have chosen
$r_{c,\text{DPD}}=1.7r_{c,\alpha\alpha}$ in our simulations.
$\zeta$ is a parameter controlling the strength of friction forces;
we set $\zeta=10$. This parameter is not crucial for the results to be
discussed \cite{Zausch2008}.
The $\mathcal N$ are Gaussian normal random variables.

The DPD equations of motion,  Eq.~\eqref{eom}, are integrated with a generalized
velocity Verlet algorithm \cite{Peters2004}. This algorithm integrates the
reversible Hamiltonian part of the equations with the velocity Verlet
scheme and then partially re-equilibrates the two-particle momenta,
ensuring a Boltzmann distribution in equilibrium. A time step of
$\delta t=5\times10^{-4}$ (in the specified units of time) is employed.
Two different neighbor
lists are implemented, one for the force calculations, and one for the
DPD thermostat. The simulation consists of $N=2N_{\text{A}}=1300$ particles
in a three-dimensional box with volume $V=L^3$ and
linear dimension $L=10\sigma_{\text{AA}}$,
corresponding to a number density $\rho=1.3$. At this density, no signs
of crystallization or phase separation were observed in the studied temperature
range from $T=5$ to $T=0.4$. A glass transition point is estimated as
$T_c\approx0.347$ according to mode-coupling theory.
In the following we focus on the equilibrated fluid at $T=0.4$.
Initial equilibration proceeded by using $\delta t=10^{-3}$, assigning new
velocities every 50 integration time steps. Equilibration was checked by
the decay of the incoherent intermediate scattering function at
a wave number corresponding to a typical interparticle separation
$q=2\pi/\sigma_{\text{AA}}$: runs were long enough to observe the decay
of the correlation function to zero at long times. A set of $200$ independently
equilibrated configurations served as initial configurations for the production
runs employing the DPD thermostat.

Shear is applied in the $x$-direction with a gradient in $y$-direction
and a positive shear rate $\dot\gamma$ initially. Planar Couette flow
is imposed by periodic Lees-Edwards boundary conditions \cite{Lees1972}:
the periodic image of a particle leaving or entering the simulation box
in $y$-direction is displaced in $x$-direction according to the strain
$\pm\dot\gamma L$.

The macroscopic response of the system to imposed shear flow is measured
through the stress tensor. In the specified coordinate system, the dominant
contribution at low shear rates is its $xy$-component, given by
the Kirkwood formula \cite{biko},
\begin{equation}
  \sigma_{xy}=\langle\hat\sigma_{xy}\rangle
  =-\frac1V\left\langle\sum_{i=1}^N\left[m_i v_{i,x}v_{i,y}
  +\sum_{j\neq i}r_{ij,x}F_{ij,y}\right]\right\rangle\,.
\end{equation}
The first term is a kinematic contribution, while the second is the
virial contribution incorporating the nonhydrodynamic forces exerted
on each particle. Angular brackets indicate canonical averages.
In the following, we restrict the discussion of the Bauschinger effect to one, rather high, exemplary shear rate, $\dot\gamma=5\times10^{-3}$.

\section{\label{sec:3}Results and discussion}

\subsection{Comparison of simulation and theory}

We begin by a discussion of the properties of the computer-simulated system under
startup of flow and in its steady state. Qualitatively, they reproduce earlier
results \cite{Zausch2008}. Here, they serve to determine the schematic model parameters prior to investigation of the Bauschinger effect.

\begin{figure}
\centerline{\includegraphics[width=.9\linewidth]{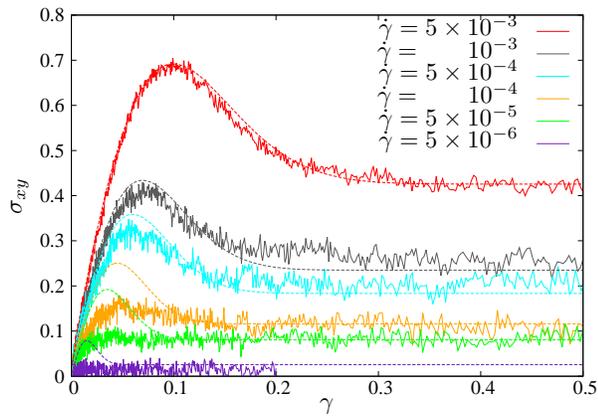}}
\caption{\label{fig:stressstrain}
  Plot of the stress-strain curves $\sigma(\gamma)$ at fixed temperature for various shear rates.
  The dashed lines show the MCT results for $\Gamma=100$, $\epsilon=-10^{-4}$, $\gamma_c=0.75$, 
  $v_{\sigma}^*=97$, and $\gamma_{\ast\ast}/\gamma_{\ast}=1.33$. $\gamma_{\ast}$ is $10^{-1}, 7 \times 10^{-2}, 
  6 \times 10^{-2}, 4.5 \times 10^{-2}, 3.5 \times 10^{-2}, 1.5 \times 10^{-2}$ for decreasing shear rates $\dot\gamma$.
  }
\end{figure}

Figure~\ref{fig:stressstrain} shows the stress--strain curves $\sigma(\gamma)$ obtained
after switch-on of steady shear at $t=0$, for various shear rates $\dot\gamma$ at one
particular temperature in the supercooled liquid. For small $\dot\gamma$, the system behaves
as a linear viscoelastic fluid, and the stress quickly reaches its steady state value,
rising monotonically from zero. As the shear rate is increased, an intermediate overshoot
in the $\sigma$-versus-$\gamma$ curve appears. Depending on $\dot\gamma$, the typical strain
reached at the position of the overshoot is a few percent; at the highest shear rates this
value is about $10\%$. It is in agreement with earlier observations \cite{Zausch2008}, and
one can argue that this typical strain corresponds to a typical size of nearest-neighbor cages
that are being broken by shear \cite{Pusey2002}.


From the large-$\gamma$ limit of the stress--strain curves, one obtains the steady-state
flow curve, $\sigma_\text{s}(\dot\gamma)$. 
The values extracted from our
simulation and the schematic-model fits exhibit shear-thinning: The stress increases  sublinearly  with the rate.
Note in Fig.~\ref{fig:stressstrain} that the values of $\sigma$ are given in simulation units, where $kT=0.4$. This
confirms that the stresses we observe are dominated by thermal motion, since
$\sigma_\text{s}=O(kT/\sigma_{\alpha\alpha}^3)$. This compares well to recent investigations of
colloidal suspensions \cite{Siebenbuerger2009,Siebenbuerger2012,Amann2012,
MarcoLauratiKJMutchNKoumakisJZauschCPAmannAndrewBSchofieldGeorgePetekidisJohnFBradyJuergenHorbach}.
The precise numerical value is found to differ
among different systems, as it somewhat depends on the 
interaction potential and mixture parameters in experimental samples.

\begin{figure}
\centerline{\includegraphics[width=.9\linewidth]{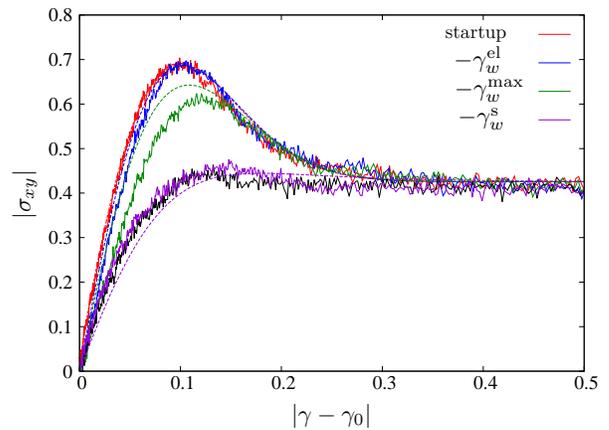}}
\caption{\label{fig:bauschinger}
  Linear plot of the
  stress-strain relation $|\sigma(|\gamma-\gamma_{0}|)|$ at fixed temperature and
  shear rate for various flow histories: starting from equilibrium (EQ,
  red line), after flow reversal in the steady state (S, purple),
  from the elastic regime (el, blue), and from the point of the
  stress overshoot (max, green). Reversal at $\gamma_w^\text{s}=0.4$ is in the steady state as seen from comparing with reversal at $\gamma_w^\text{s}=75$ (black data). Smooth lines result from the schematic model.}
\end{figure}

The solid lines in Fig.~\ref{fig:stressstrain} are fits with the schematic MCT model
including the time-dependent vertex for the Green-Kubo relation. These fits were optimized
to reproduce the shape of the overshoot at the largest shear rate (for which the Bauschinger
effect will be discussed below). Further adjustments to improve agreement at the lower
shear rates may well be possible, but note that the form of the overshoot is dictated by the
empirical choice of $v_\sigma(t,t')$.

The parameters of the schematic MCT model can be determined by fitting the
simulation data in startup shear. Using a fit procedure similar to
the one outlined in Ref.~[\onlinecite{Amann2012}] we arrive at the values listed in the caption of Fig.~\ref{fig:stressstrain}.
The parameters $\Gamma$ and $\epsilon$ determine the short- and long-time relaxation scales
$\tau_0$ and $\tau$ in the model. They are fixed by a comparison of the quiescent
equilibrium correlation functions at a typical intermediate wave number corresponding
to the first peak of the static structure factor. Linear-response fluctuations also
allow to determine $v_\sigma^*$, corresponding to the plateau modulus $G_\infty$ and
hence setting the units of stresses in the model. The parameters $\gamma_c$,
$\gamma^*$, and $\gamma^{**}$ are then tuned to the startup rheology data,
taking the shear rate $\dot\gamma$ nominally from the simulation.

Once the parameters given in the caption of Fig.~\ref{fig:stressstrain} have been fixed, the predictions of the schematic
model regarding the modification of the stress--strain curve under flow reversal are
parameter-free.
%
%
%
%
%
%
%
%
%
%
%

We now turn to the discussion of the Bauschinger effect.
Figure~\ref{fig:bauschinger} shows the main result of the computer simulation,
viz.\ the stress--strain curves in the $|\sigma|$-versus-$|\gamma-\gamma_0|$ representation,
i.e., shifted and inverted such that they all start from a stress-free state
and display the behavior as a function of the additional strain imposed on this configuration.
Curves for various $\gamma_w$ are shown, as discussed above. They all agree at large
strains by construction, confirming that the steady-state stress does not depend
on the shear history.
Comparing first the startup curve, $\gamma_w=0$, with the one for reversal in the
steady-state, $\gamma_w=\gamma_w^\text{s}$, the Bauschinger effect becomes most apparent.
The two curves differ in two main aspects: the linear slope at small deformations
is lower, and the overshoot is gone in the curve corresponding to oppositely straining the (steadily)
pre-sheared configuration. This holds for shear-reversal throughout the stationary state, as shown by the two curves for $\gamma_w^\text{s}=0.4$ and $\gamma_w^\text{s}=75$ included in Fig.~\ref{fig:bauschinger}.  Both these differences arise because the system undergoes
plastic deformations during pre-shear: confining the pre-shear to small strains,
$\gamma_w=\gamma_w^\text{el}$, so that flow reversal takes place inside the initial
elastic-deformation regime, the overshoot is maintained almost unchanged, as is the
effective elastic coefficient extracted from the initial rise of the stress--strain curve.
The cross-over between elasticity-dominated and plasticity-dominated pre-strain occurs
gradually, as the curve for $\gamma_w=\gamma_w^\text{max}$ exemplifies.

The solid lines in Fig.~\ref{fig:bauschinger} represent the schematic-MCT model calculations.
The line corresponding to $\gamma_w=0$ is a result of the fitting procedure
described above. All the other theory curves then follow from the structure of the
ITT-MCT equations. They describe the computer-simulation data extremely well, and
capture both the decrease of the overshoot and the decrease of the effective elastic
coefficient.

\begin{figure}
\centerline{\includegraphics[width=.9\linewidth]{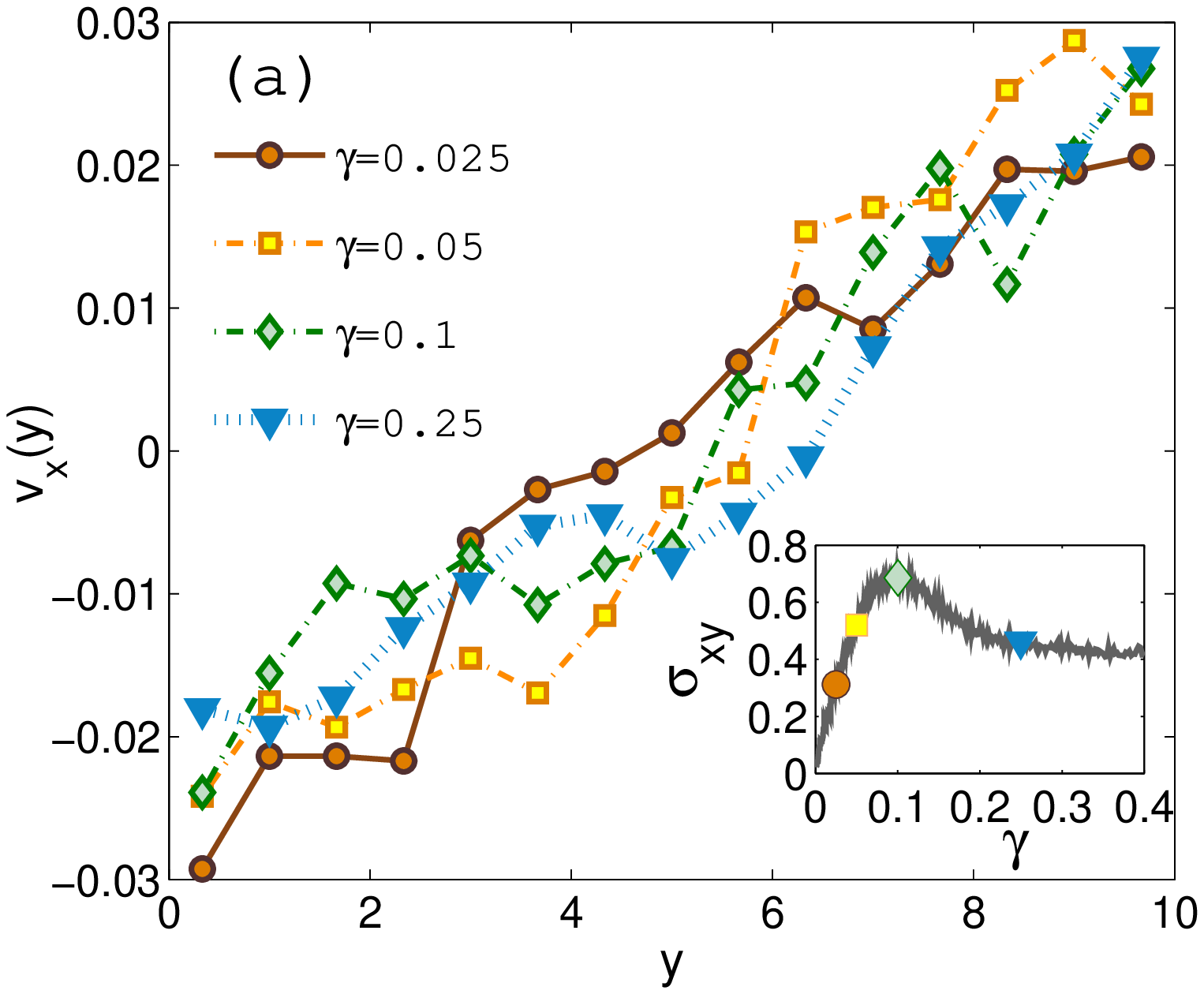}}
\centerline{\includegraphics[width=.9\linewidth]{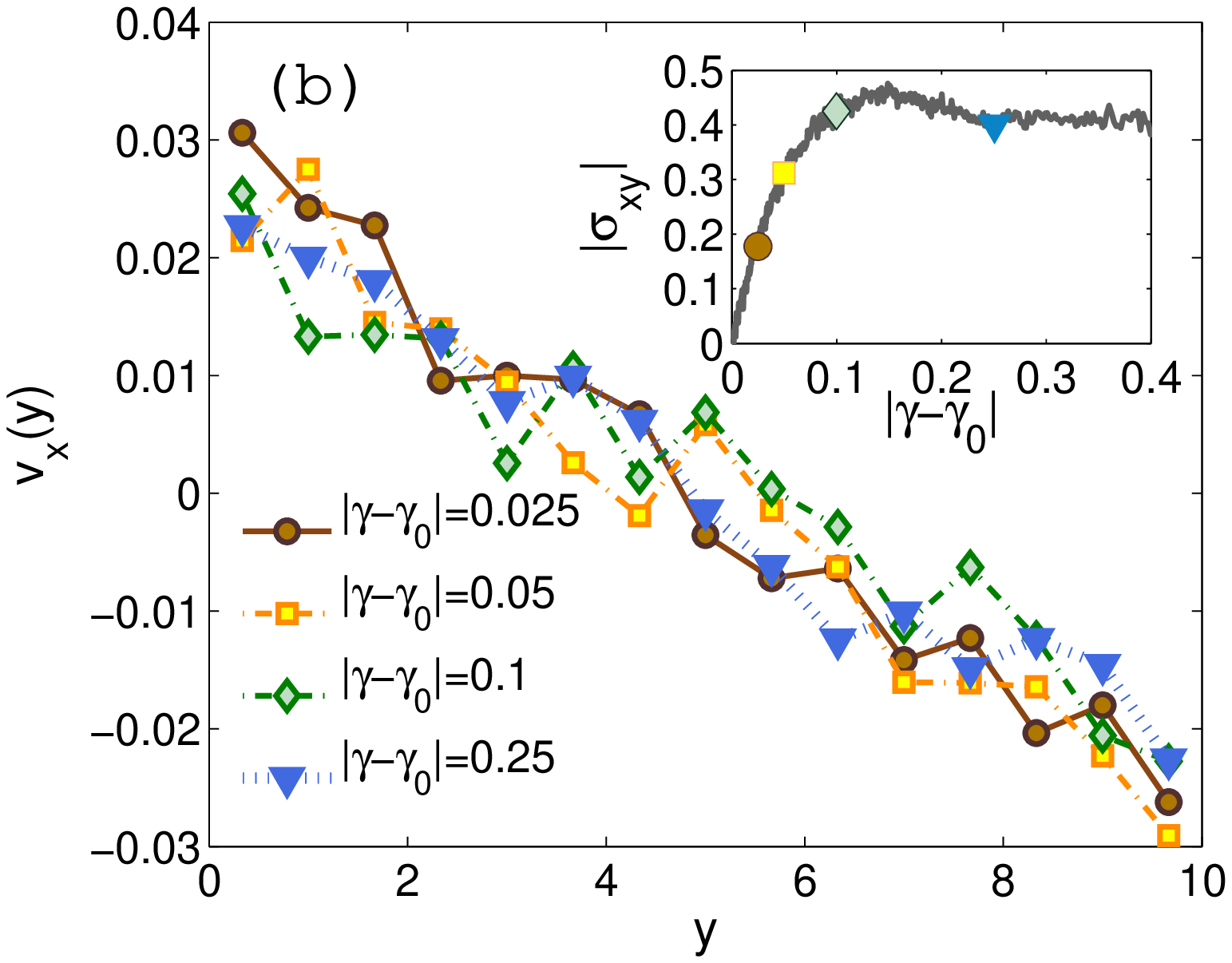}}
\caption{\label{fig:profiles}
  Velocity profiles $v_x(y)$ as a function of position along the
  gradient direction $y$ for various times after application of a
  new flow direction.
  (a) Starting from a quiescent configuration (EQ), shear in $+x$ direction.
  (b) Starting from the resulting steady state (S) at $\gamma^s_w=0.4$, shear in $-x$ direction.
  Insets mark the points for which profiles are shown along the
  stress-strain curves. A total $100$ independent configurations are averaged 
  within a small time window to obtain the depicted graphics.
}
\end{figure}

Before continuing the discussion, it is worth pointing out that none of the
features we discuss appear to be explained by, or connected to shear banding
or other inhomogeneous flow effects. To demonstrate this, we show in
Fig.~\ref{fig:profiles} the velocity profiles obtained from the simulation
at various instants in time, both under forward shear and the subsequently
reversed shear. To improve statistics, the instantaneous velocities were
averaged over all the configurations and over a small time window of
width $\Delta t=0.1$.
Insets in the figure mark the points for which the
velocity profiles $v_x(y)$ are shown; they all follow the linear behavior
$v_x(y)=\dot\gamma y$
expected from homogeneous simple shear, to within statistical noise.
Similar results have also been obtained under startup shear \cite{Zausch2008}.
It appears that the small time needed to establish a linear velocity profile
under Lees-Edwards boundary conditions is not relevant for our discussion.
In Fig.~\ref{fig:profiles} only the case for flow reversal in the steady
state is shown, but qualitatively the same result is found for the
velocity profiles obtained after switching the shear direction at the
smaller $\gamma_w^\text{max}$ or $\gamma_w^\text{el}$.
Stress overshoot phenomena have been linked to shear banding \cite{Moorcroft2011},
but it appears that this mechanism is not relevant for our system.
Indeed, our results are compatible with a recent study suggesting a
density-dependent critical P\'eclet number for the appearance of shear
bands \cite{Besseling2010}.

\subsection{Interpretation in a generalized Maxwell model}

To understand the history dependence of the stress--strain curves, let us split the integral in Eq.~\eqref{eqn:greenkubo}
into two parts. For $t>t_w$, write
\begin{equation}
  \sigma(t)=\sigma_\text{I}(t)-\sigma_\text{II}(t)
\end{equation}
with the summands
\begin{subequations}
\begin{align}
  \sigma_\text{I}(t)&=\dot\gamma\int_0^{t_w}\text dt'\,G(t,t')\,, \label{sigi1}\\
  \sigma_\text{II}(t)&=\dot\gamma\int_{t_w}^t\text dt'\,G(t,t')=\dot\gamma\int_{t_w}^t\text dt'\,G(t-t')\,.\label{sigi2}
\end{align}
\end{subequations}
In the last term, $\sigma_\text{II}(t)$, the condition $t_w<t'<t$ is fulfilled, so that the
correlation and strain functions entering the integral are dependent on one-time only and describe the transient between an equilibrium and
a stationary state. Hence,
this contribution is trivially related to the startup curve,
$-\sigma_\text{II}(t+t_w)=\sigma_\text{start}(t)$; see below in Sect.~\ref{sec:4c} for details.

To simplify the discussion, let us consider an ad-hoc generalization of the Maxwell model
of viscoelastic liquids to shear-thinning fluids. Keeping the form of the Green-Kubo equation, we
approximate the generalized shear modulus, Eq.~\eqref{eqn:gschem}, by
\begin{equation}\label{maxwell}
  G(t,t')\approx G_M(t,t')=v_\sigma(t,t')e^{-(t-t')|\dot\gamma(t')|/\tilde\gamma_c}\,.
\end{equation}
For the case of constant shear and constant $v_\sigma$,
this form shows the same qualitative features as the shear-molten
glass in schematic MCT \cite{fuca}. To generalize this nonlinear Maxwell model to
unsteady flows is neither trivial nor unique, but assuming the rate of decorrelation to be set
by the instantaneous shear rate, the above ansatz appears plausible, if somewhat crude \cite{Siebenbuerger2012}. In the present case it even simplifies further because $|\dot\gamma(t')|=\dot\gamma$. 
The integrals determining
$\sigma(t)$ and its two contributions $\sigma_\text{I}$ and $\sigma_\text{II}$
can then be solved analytically when inserting the form of $v_\sigma(t,t')$ given above.
In order to quantitatively match the startup stress--strain curve of the generalized Maxwell model
with the schematic model, we set $\tilde\gamma_c=\gamma_c/3$.

\begin{figure}
\centerline{\includegraphics[width=.9\linewidth]{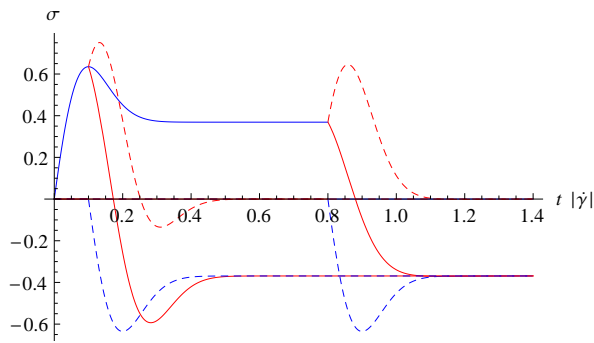}}
\caption{\label{fig:maxbausch}
  Bauschinger effect as illustrated by a generalized Maxwell model (see text).
  Solid lines are $\sigma(t)$ as a function of normalized time, $|\dot\gamma|t$,
  for the case $\gamma_w=0.1$ corresponding to $\gamma_w^\text{max}$ and
  $\gamma_w=0.8$ corresponding to $\gamma_w^\text{s}$. Dashed lines display
  the contributions $\sigma_\text{I}(t)$ (upper, red) and
  $\sigma_\text{II}(t)$ (lower, blue).
}
\end{figure}

The response for large $t_w$ can now be understood, recalling that the integral appearing in
$\sigma_\text{I}(t)$ is dominated by $t'\approx t_w$, where the integrand is still
non-vanishing. The accumulated strain entering the correlation functions can then be written
as $\gamma_{tt'}=\dot\gamma(2t_w-(t+t'))\approx\dot\gamma(t-t_w)$. Hence, $\sigma_\text{I}$
becomes a function of $t-t_w$ only, and the precise time of switching the flow becomes irrelevant
(as it takes place in the steady state after startup).
This is confirmed by numerical evaluation and also in the simulation, by varying
$\gamma_w^\text{s}$.
At $t=t_w$, $\sigma_\text{I}$ is nothing but the value of the startup curve; as $t\to\infty$,
this contribution vanishes since the functions in the integrand will generally decay.
Furthermore, at $t\approx t_w+\gamma_c/\dot\gamma$, a maximum occurs as the relevant
strain $|\gamma_{tt'}|\approx\gamma_c$ when the integral is dominated by large $t'$.
This maximum cancels out the corresponding minimum from $\sigma_\text{II}$, so that
the stress overshoot is generically expected to vanish. This is shown in
Fig.~\ref{fig:maxbausch} for the case $\gamma_w=0.8$.
For $t_w\approx\gamma_w^\text{max}/\dot\gamma$,
however, the integral determining $\sigma_\text{I}$ will have two relevant contributions,
namely $t'\approx t_w$ as well as $t'\approx0$, since the relevant strain at both
these times equals $\gamma_c$. Hence, $\sigma_\text{I}(t)$ in this
case displays first a maximum and then a minimum, as exemplified in Fig.~\ref{fig:maxbausch}
for the case $\gamma_w=0.1$. The minimum is responsible for $\sigma_\text{I}(t)$  not to cancel the stress overshoot contained in $\sigma_\text{II}(t)$.
Obviously, as $t_w\to0$, the contribution $\sigma_\text{I}(t)$ vanishes, so that
perfect elastic recovery is obtained in this limit. 
While the generalized Maxwell model simplifies the interpretation, the next sections show that broadly equivalent curves follow from the present schematic model, which provides a fundamental basis for this simplification. It exhibits a two step process, where the final decay recovers the properties of Eq.~\eqref{maxwell}.

\begin{figure}
{\includegraphics[width=.9\linewidth]{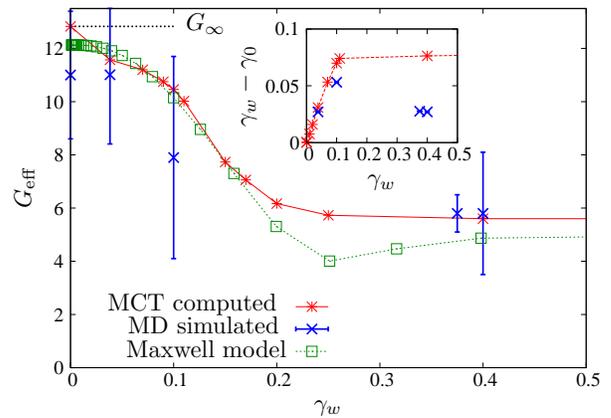}}
\caption{\label{fig:geffrelover}
  Linear stress responses $G_{\text{eff}}=d\sigma/d\gamma|_{|\gamma-\gamma_{0}|=0.025}$ from simulation, fits of the schematic model, and Maxwell model as labeled   
for various waiting strains $\gamma_{w}$.  The inset gives the strain $\gamma_{w}-\gamma_{0}$  required for the stress to
reduce to zero after shear reversal. To include the simulation point for $\gamma_w^s=75$, this value has been divided by 200, using that the precise value of $\gamma_w$ is irrelevant in the steady state.}
\end{figure}

We now turn to a discussion of the effective shear moduli, obtained as
$G_{\text{eff}}=d\sigma/d\gamma|_{|\gamma-\gamma_{0}|=0.025}$. They measure the effective elasticity remaining until the shear-driven relaxation sets in.  
Figure~\ref{fig:geffrelover} displays the results from the schematic MCT model (red and blue stars)
together with the values extracted from the simulation data (blue crosses). The agreement is
quite satisfactory, noting that only the initial value at $\gamma_w=0$ is a result of a
fitting procedure. One observes a marked decrease of $G_\text{eff}$ around $\gamma_w\approx0.1$,
the position of the overshoot in the startup curve. Hence, pre-shear indeed softens the
material, but only by plastic deformation.
Also shown in Fig.~\ref{fig:geffrelover} are the results from the generalized Maxwell model;
qualitatively the same trend is seen. In the Maxwell model, we could not find a closed expression to determine
$\gamma_0$; but the effective shear modulus defined as $\gamma\to\gamma_w$ for $t\to t_w+0$ can
be evaluated analytically. It again follows the trend seen in the figure, albeit with an even
lower limiting value at large $\gamma_w$.
%
%

A missing information
on the simulated stress-strain curve remains $\gamma_0$, the strain value where the stress-free state is achieved after strain reversal. 
Its qualitative behavior can be understood easily. Assuming that $G_\text{eff}$ is unchanged for
small $\gamma_w$,  the relations $\sigma(t_w)=G_\text{eff}\gamma_w$
and $0=\sigma(t_0)=\sigma(t_w)-G_\text{eff}(\gamma_0-\gamma_w)$ hold. As a result, $\gamma_0=2\gamma_w$.
Linearity holds even if $G_\text{eff}$ changes, but the prefactor is somewhat different, as observed in the simulations.
For large $\gamma_w$, $0=\sigma(t_0)=\sigma_\text{ss}-G_\text{eff}(\gamma_0-\gamma_w)$,
and the constant value in Fig.~\ref{fig:geffrelover} is indeed roughly $\sigma_\text{ss}/G_\text{eff}\approx0.4/6\approx0.07$. 

\subsection{Discussion of the history dependence in the schematic model}\label{sec:4c}

\begin{figure}
\centerline{\includegraphics[width=.9\linewidth]{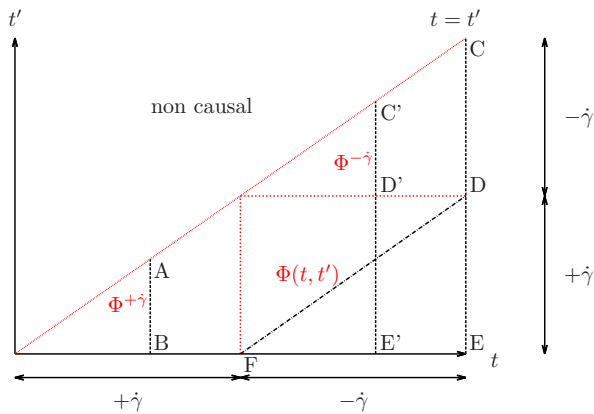}}
\caption{\label{fig:timeplane}
  View on the $t-t'$-plane, with the three different time regimes for the density correlator $\Phi$.
  As the $h$-function depends only on the absolute value of $\dot\gamma$, the one-time-solutions
  $\Phi^{(+\dot\gamma)}$ and $\Phi^{(-\dot\gamma)}$ are equal.
  }
\end{figure}

To understand the history dependence of the stress--strain curves within the schematic ITT-MCT model requires to analyze the time-dependence of the correlator
which encodes plastic deformations, and to combine it with the one of the elastic coefficient $v_\sigma$, which encodes
strain-dependent anelasticity. 
The structure of Eq.~\eqref{eq:eom} and the time-dependence of the accumulated strain
suggests to split the integration domain into three regimes, as shown in Fig.~\ref{fig:timeplane}. It sketches the corresponding integration paths in the $t-t'$--plane leading to the stresses shown in Fig.~\ref{fig:hysteresis} using Eq~\eqref{eqn:gschem}. 
Immediately, the strain follows from Eq.~\eqref{shearrate} (remembering that $t'<t$):
\begin{equation}
  \gamma(t,t')=\begin{cases} \dot\gamma \left(t-t'\right) & \text{for $0<t'<t< t_w$,}\\
\dot\gamma \left( 2 t_w-(t+t')\right) & \text{for $0<t'<t_w<t$,}\\
  -\dot\gamma \left(t-t'\right) & \text{for $t_w<t'$,}\end{cases}\label{strain}
\end{equation}
which identifies the three regimes. An integration for $t',t<t_w$ gives the startup stress--strain curve, the integration  for $t'<t_w<t$ gives the stress contribution $\sigma_I$ in Eq.~\eqref{sigi1}, and the  integration  for $t_w<t',t$ the  contribution $\sigma_{II}$ in Eq.~\eqref{sigi2}.
Recall that
the correlators appearing in the MCT model are transient correlation functions, i.e., they
are formed with the equilibrium distribution function always, although the full nonequilibrium
dynamics appears in their propagator. A correlation function $\Phi(t,t')$ hence contains
only information on the flow history in the interval $[t',t]$. In particular, the correlation
functions defined for $0<t'<t<t_w$ are given by the startup transient solution $\Phi^{(\dot\gamma)}$
for $\dot\gamma(t)=+\dot\gamma$ of the model;
the same holds true for those defined for $t_w<t'<t$, since there the shear rate is constant and the direction of shear cannot enter by symmetry. Because the strain in these two time regimes depends on the time difference only (see Eq.~\eqref{strain}), the correlator $\Phi^{(\dot\gamma)}(t-t')$
also depends on the time difference only. This greatly reduces the complexity of Eq.~\eqref{eq:eom}
and allows the use of well-established integration schemes to solve
\begin{align}
 \dot\Phi^{(\dot\gamma)}(s) + \Gamma\Phi^{(\dot\gamma)}(s)
  + \Gamma\int\limits_{0}^{s}\!\text{d}s'\, m^{(\dot\gamma)}(s-s')\dot\Phi^{(\dot\gamma)}(s') =0,\nonumber\\ 
m^{(\dot\gamma)}(s) = h(\gamma(s)) \left[ v_1 \Phi^{(\dot\gamma)}(s) + v_2 \Phi^{(\dot\gamma)}(s)^2\right]\; .
\label{einzeitenEOM}
\end{align}
Here, $s=t-t'$  abbreviates the time difference.
The correlation function for $t'<t_w<t$ however remains fully two-time dependent, as here the strain depends on $t+t'$. It is
convenient to split the integral appearing in Eq.~\eqref{eq:eom} explicitly at $t''=t_w$,
\begin{align}
 \partial_{t}\Phi(t,t') + \Gamma\Phi(t,t') +
\Gamma\int\limits_{t'}^{t_{w}}\!\text{d}t''\,
m(t,t'',t')\dot\Phi^{(\dot\gamma)}(t''-t')\nonumber\\
 + \Gamma\int\limits_{t_{w}}^{t}\!\text{d}t''\,
m^{(\dot\gamma)}(t-t'')\partial_{t''}\Phi(t'',t_{w})=0\;,
\label{zweizeitenEOMsplit}
\end{align}
Here, $m^{(\dot\gamma)}$ is the same memory kernel also appearing in Eq.~\eqref{einzeitenEOM}.
Hence, in this formulation, one-time correlators or memory kernels previously calculated enter most succinctly.
Supplied with suitable initial conditions, Eq.~\eqref{zweizeitenEOMsplit} is a convenient way
to deal with instantaneous shear-rate switches numerically \cite{Voigtmann2012}.

\begin{figure}
\centerline{\includegraphics[width=.9\linewidth]{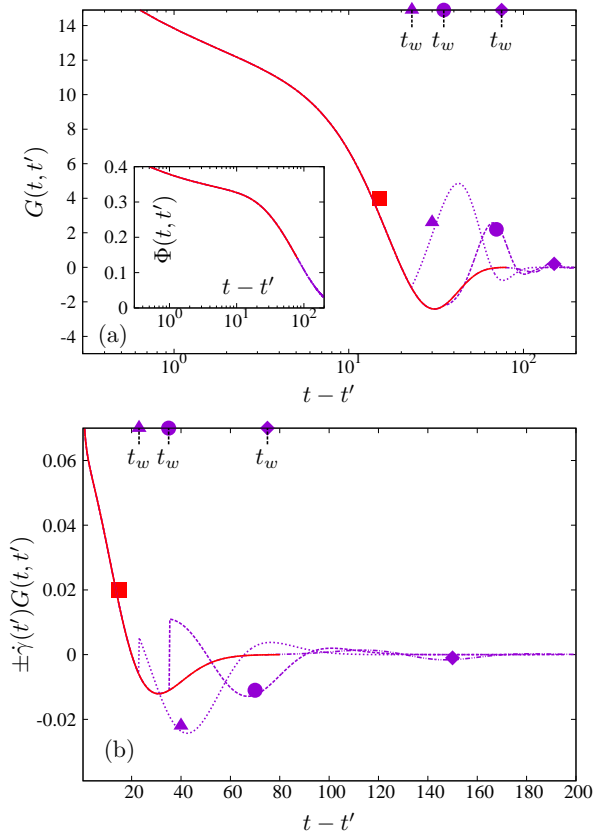}}
\caption{\label{fig:3erpanel_for_reversal_in_steadystate}
(a) Shear moduli $G(t,t')$ after shear reversal at $\gamma^s_w$ in the steady state, shown
for three different fixed times $t$ (respectively strains $\gamma$), as functions of $t-t'$.
Curves marked with violet triangle, disc, and diamond symbols correspond to points marked with the
same symbol in Fig.~\ref{fig:hysteresis}.
For times $t-t'<t_w$, all curves collapse by construction onto the startup result $G(t,0)$
shown in red (marked with a square).
The inset shows the corresponding correlators  $\Phi (t,t')$ (they overlap on the resolution of the figure).
(b) Corresponding stress integrands $\pm\dot\gamma(t')G(t,t')$ on a linear $t-t'$ axis.
Curves after shear reversal which would be negative for $t-t'\to0$ are mirrored to positive initial decay, 
where they overlap with the startup curve. }
\end{figure}

\begin{figure}
\centerline{\includegraphics[width=.9\linewidth]{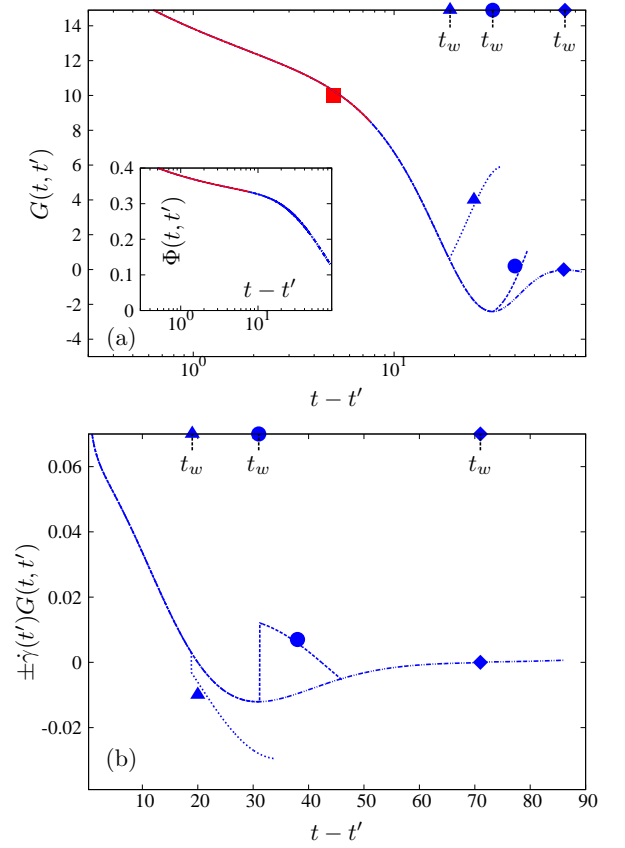}}
\caption{\label{fig:3erpanel_for_reversal_in_elastic_regime}
Like Fig.~\ref{fig:3erpanel_for_reversal_in_steadystate}, but now for shear reversal at $\gamma^\text{el}_w$ in the elastic regime. The curves end at a time $t-t'=\gamma^\text{el}_w/\dot\gamma$, where the memory has not decayed to zero yet. This is as expected, because the system still responds elastic like before the early reversal at $\gamma^\text{el}_w$; see the blue curve in Fig.~\ref{fig:hysteresis}.
}
\end{figure}

Evaluation of a stress $\sigma(\gamma=\int_0^tdt'\dot\gamma(t'))$ requires integrations of the transient shear modulus along lines at constant $t$ in the plane shown in Fig.~\ref{fig:timeplane}. The startup curve is obtained along lines of the type from A to B before the shear reversal, where all functions depend on $t-t'$. Figures \ref{fig:3erpanel_for_reversal_in_steadystate} and \ref{fig:3erpanel_for_reversal_in_elastic_regime} include the transient density correlators, the transient shear moduli and the integrands of Eq.~\eqref{eqn:greenkubo} for this case. As discussed in detail in Refs.~[\onlinecite{Zausch}, \onlinecite{Amann2012},  \onlinecite{MarcoLauratiKJMutchNKoumakisJZauschCPAmannAndrewBSchofieldGeorgePetekidisJohnFBradyJuergenHorbach}], 
the linear regime in the stress--strain curve is connected to the plateau on intermediate times in $\Phi(t)$ and $G(t)$, and the stress-overshoot to the negative dip in $G(t)$ at late times. Reversing the direction of the shear at a late time when the stationary state was reached at $\gamma^s_w$, correlators and shear moduli change as shown in Fig.~\ref{fig:3erpanel_for_reversal_in_steadystate}. Three different final states are chosen along the reversing stress-strain curve: One in the elastic regime, one in the regime where the stress-maximum is now missing, and one in the late steady regime, where the  steady state is reached again. Obviously, the last curves agree with the startup curves (except for the trivial minus sign in the integrand), as the same transient into a steady state is probed. Here the integration path from C to D, where the shear rate takes one definite (negative) value, dominates and the time region, where the shear rate was different, affects the very final transients, only. Moving the final time $t$ to correspond to a strain $\gamma$ of around 10\%, where the maximum in the startup curve appears, the transient shear modulus is affected in the negative region. There the strong variation in $v_\sigma(t,t')$ of Eq.~\eqref{eqn:ammanprefactor}   around $\gamma^*$ becomes important. Along the integration path from D' to E', the memory of the $+\dot\gamma$ shear enters, which causes the sign change in $G(t,t')$; see the curve marked with a circle in panel (a) of Fig.~\ref{fig:3erpanel_for_reversal_in_steadystate}. The integrand of the generalized Green-Kubo relation  Eq.~\eqref{eqn:greenkubo} varies around zero, see panel (b) of Fig.~\ref{fig:3erpanel_for_reversal_in_steadystate}, in the region where the stress overshoot arose. This leads to a cancellation in the integral, and an absence of the maximum in the stress.  Shifting the final time $t$ to earlier values in the region, where the startup stress varies linearly with strain, causes the sign change in $G(t,t')$ to move to earlier times $t'$ as well;  see the curve marked with a triangle in panel (a) of Fig.~\ref{fig:3erpanel_for_reversal_in_steadystate}. A stronger memory of the shear rate with opposite sign remains. As this arises especially during the final relaxation, the region in $G(t,t')$ where stresses are anticorrelated, the overall integral over the transient shear modulus is smaller. As $t$, respectively $\gamma$, lies in the region, where the linear increase in the stress occurs, this explains the softening of the effective elastic constant $G_\text{eff}$; see the linear slope indicated in Fig.~\ref{fig:hysteresis} at the violet triangle.  Importantly, in the ITT-MCT approach it arises not by a weakening of the plateau in the transient correlation functions, rather from the aftereffect of the previously stored stresses accumulated during the flow with opposite (positive) shear rate. It thus originates from the final relaxation and does not hold in the proper limit of infinitesimal strain. We chose $\gamma=\gamma^*/4$ to measure $G_{\rm eff}$. 

The equivalent discussion holds for reversing the flow during the elastic regime of the startup stress-strain curve. This is indicated by a  blue curve in Fig.~\ref{fig:hysteresis}, and the corresponding transient functions are shown in Fig.~\ref{fig:3erpanel_for_reversal_in_elastic_regime}. Again a triangle, circle, and diamond marks final states on the reversing stress--strain curve in the elastic, overshoot, and steady regime.  The integration paths in the $t,t'$-plane of Fig.~\ref{fig:timeplane} do not change qualitatively, but the time $t_w$ of switching is now so short that the startup correlators along the line A to B have not decayed to zero. The system responds elastically, and the correlator $\Phi^{(\dot\gamma)}$ and the modulus  $G^{(\dot\gamma)}$
are still of the order of the plateau value, $f$ respectively $G_\infty$. The latter value determines the linear increase of the startup stress-strain curve. The complete discussion of the changes in $G(t,t')$ done in respect to Fig.~\ref{fig:3erpanel_for_reversal_in_steadystate} carries over to  
Fig.~\ref{fig:3erpanel_for_reversal_in_elastic_regime} with the sole difference that the two-time region D to E (and D' to E') dominates at first. By continuity, it reproduces the stress $\sigma(\gamma^\text{el}_w)$, where the flow reversal took place. Because the strain increases with $t+t'$ in this region according to Eq.~\eqref{strain}, this contribution decays rapidly along the reversing stress-strain curve. It may be neglected beyond $\gamma_-\gamma_w$, where the initial elastic stress is destroyed. The remaining contribution along the lines C to D or C' to D' is equivalent to the startup flow along A to B, and thus after shear rate reversal in the elastic regime, the stress-strain curve agrees closely with the initial startup curve.  

\begin{figure}
{\includegraphics[width=.9\linewidth]{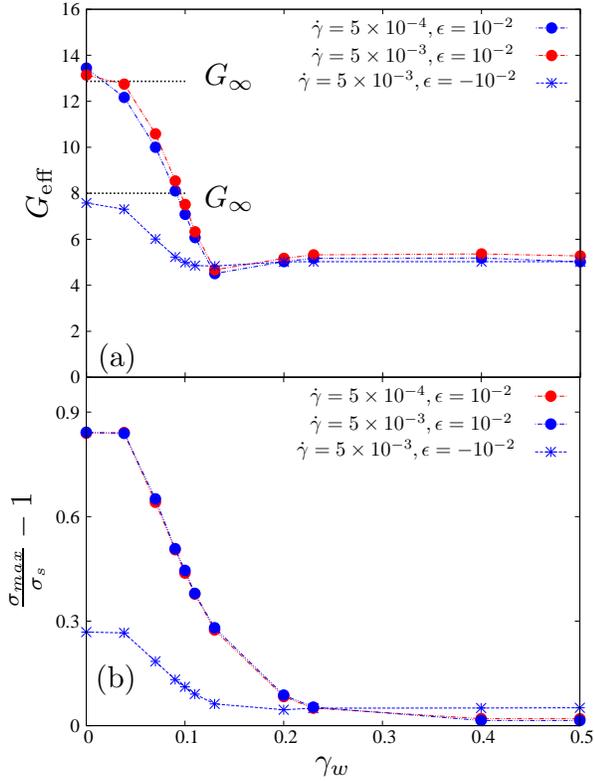}}
\caption{\label{fig:geffrelovertheo}
In panel (a), linear stress response $G_{\text{eff}}=d\sigma/d\gamma|_{|\gamma-\gamma_{0}|=0.025}$ as function of waiting strain  $\gamma_{w}$  in a fluid and a glass state; for the latter, two different shear rates are shown. The quiescent elastic shear constant $G_\infty$ is observed in startup flow. \\
In panel (b), the corresponding curves are given for the
  relative stress overshoots $\sigma_{max}/\sigma_{s}-1$.}
\end{figure}

The preceding discussion of the Bauschinger effect at two typical  waiting or reversal times $t_w$ corresponding to strains in the elastic and steady regime enables one to rationalize the stress-strain curves for all pre-shears $\gamma_w$. Figure \ref{fig:geffrelovertheo} summarizes the main findings for two different temperatures close to the glass transition in terms of the effective shear constant $G_\text{eff}$ and the relative amplitude of the stress-overshoot  $\sigma_{max}/\sigma_{s}-1$. Both change from their quiescent values as soon as $\gamma_w$ becomes of the order of $\gamma^*$, where the transient shear modulus captures anticorrelated stress fluctuations. The variation of $G_\text{eff}$ saturates when $\gamma_w>\gamma^{**}$, because then stress correlations have become negligible; then also the overshoot  $\sigma_{max}/\sigma_{s}-1$ is gone. Increasing the static correlations in the system, mimicked by $\epsilon$ in the schematic model, increases the elasticity. Because bare Peclet number Pe$_0\ll1$ are considered, no qualitative difference holds for a liquid and a glassy state. Decreasing the shear rate in the glass, the discussed effects remain and just become somewhat smaller.  In a fluid state, the linear response regime would be approached as soon as the dressed Peclet or Weissenberg number Pe$=\dot\gamma\tau$ becomes of order unity. This situation where no overshoot arises \cite{Amann2012} and thus flow reversal only mildly affects the stress-strain curve is not shown here.

\subsection{Simulation results on average particle motion}

\begin{figure}
\centerline{\includegraphics[width=.9\linewidth]{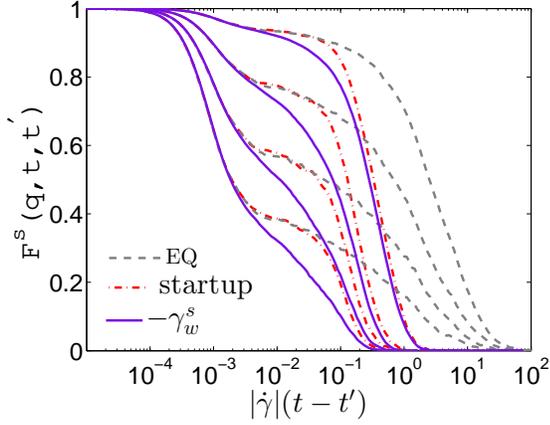}}
\caption{\label{fig:fsqt_shear}Simulation results of the self intermediate scattering functions 
$F_s(q, t)$ for the B-species in the two
 directions perpendicular to shear during startup
 and after shear reversal in the stationary state, for wave vectors $q = [\pi/\sigma_{AA},
{2\pi}/\sigma_{AA},{3\pi}/\sigma_{AA},
{4\pi}/\sigma_{AA}]$ in descending order. A total 200 independent
simulational runs are averaged to obtain these graphics. The corresponding equilibrium functions are included (dashed lines, label EQ), shifted to coincide at short times. }
\end{figure}

To relate the macroscopic response discussed so far to the microscopic dynamics, we turn to
a discussion of simulation results for the density correlation functions under shear.
Figure~\ref{fig:fsqt_shear} shows as exemplary cases the self-intermediate scattering functions
(tagged-particle density-correlation functions) $F^s(q,t,t')$ in the quiescent equilibrium ($t'\to-\infty$, dashed),
under startup flow ($t'=0$, dash-dotted), and after flow reversal from the steady state ($t'=\gamma_w^\text{s}/\dot\gamma$ using $\gamma_w^s=0.4$,
solid lines). Each curve has been sampled by averaging over 200 initial configurations, and we checked that stationarity is achieved comparing with reversal at $\gamma_w^s=75$. To
avoid a discussion of shear-advection effects, wave vectors were chosen with zero component
in the shear direction.
The equilibrium curves exemplify the typical two-step relaxation found in glass-forming
liquids: after an initial fast relaxation to an intermediate plateau, the final decay to zero
is characterized by a large time scale $\tau\approx10^3$. This final decay is much more stretched
than exponential relaxation. The plateau value depends on the wave
number $q$, and for the correlation functions characterizing tagged-particle motion generically
decreases with increasing $q$. Steady shear accelerates the dynamics, and induces relaxation
from the plateau on a time scale set by $1/\dot\gamma$ (with a $q$-dependent prefactor). This
relaxation is much closer to exponential, and, for the transient correlation function shown here,
even slightly compressed. These qualitative features confirm earlier observations
\cite{Zausch2008,bewescpo,Krueger2011}.
By symmetry, the same is true for the correlation functions probing the reversed-flow regime
(solid lines in the figure). However, as in the macroscopic response, a pronounced history
dependence is seen: the intermediate plateau is much less pronounced, and the final shear-induced
decay shows much more pronounced stretching, in particular at large $q$.

In discussing these dynamics, one has to emphasize the difference of these
measurable quantities to the transient correlation functions used in MCT.
The theory defines correlation functions as averages over the equilibrium distribution; they
have the advantage to be better suited for subsequent approximations. In the simulation, such
correlation functions can only be measured (beyond the quiescent equilibrium) under startup
of steady shear from an equilibrated configuration. This is the case shown in
Fig.~\ref{fig:fsqt_shear} by the dash-dotted lines, which can be compared  qualitatively  ---  in view of the lack of a $q$-dependence  of the latter ---  to correlators from the schematic model included in the inset of Fig.~\ref{fig:3erpanel_for_reversal_in_steadystate}. 
For the case of flow reversal
(solid lines), the simulation implies averaging over the steady-state distribution
corresponding to forward flow, and not the quiescent one. We observe that this quantity shows changes already in
the relaxation around the plateau, and not just for features of the final relaxation.

\begin{figure}
\centerline{\includegraphics[width=\linewidth]{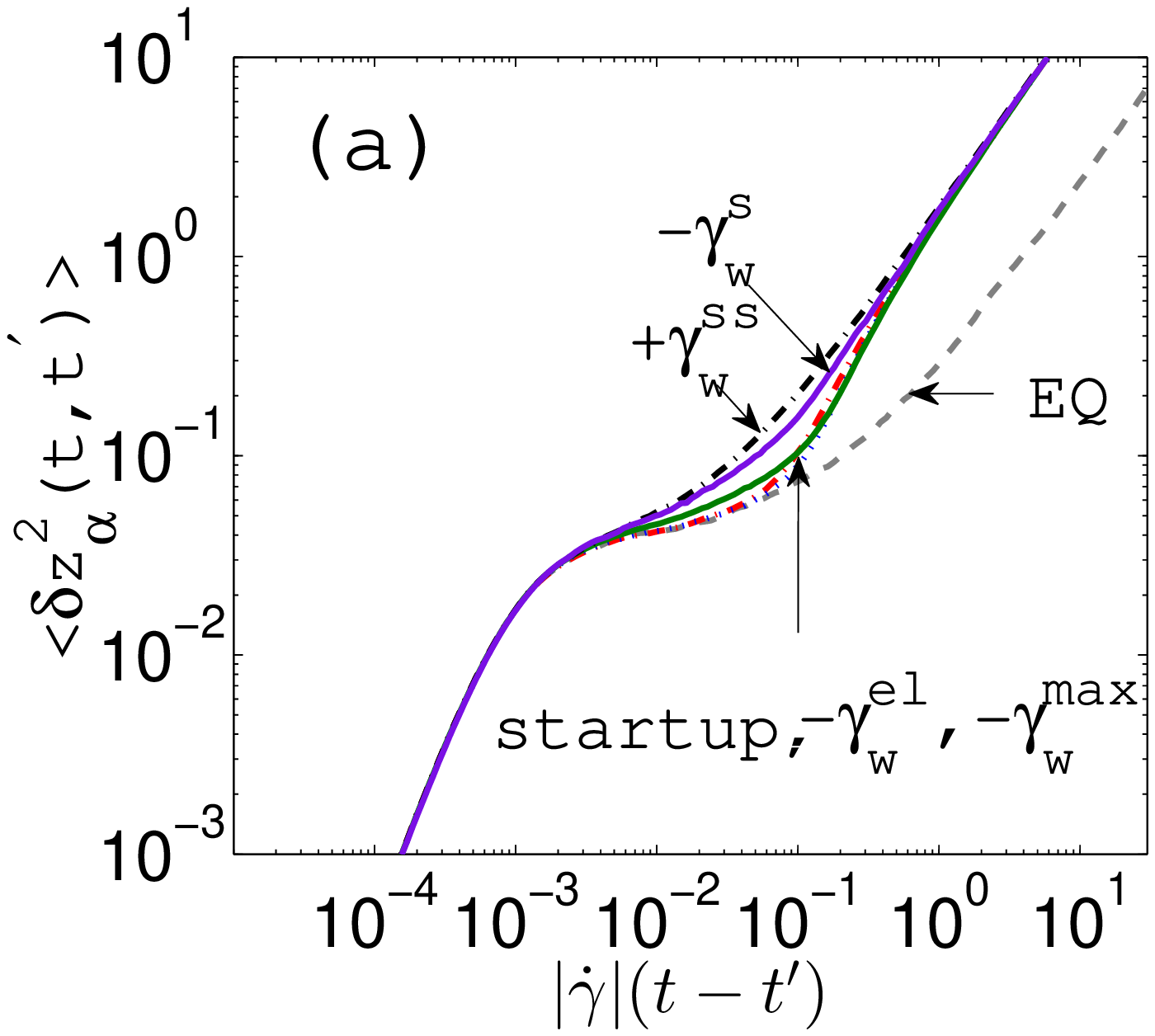}}
\centerline{\includegraphics[width=\linewidth]{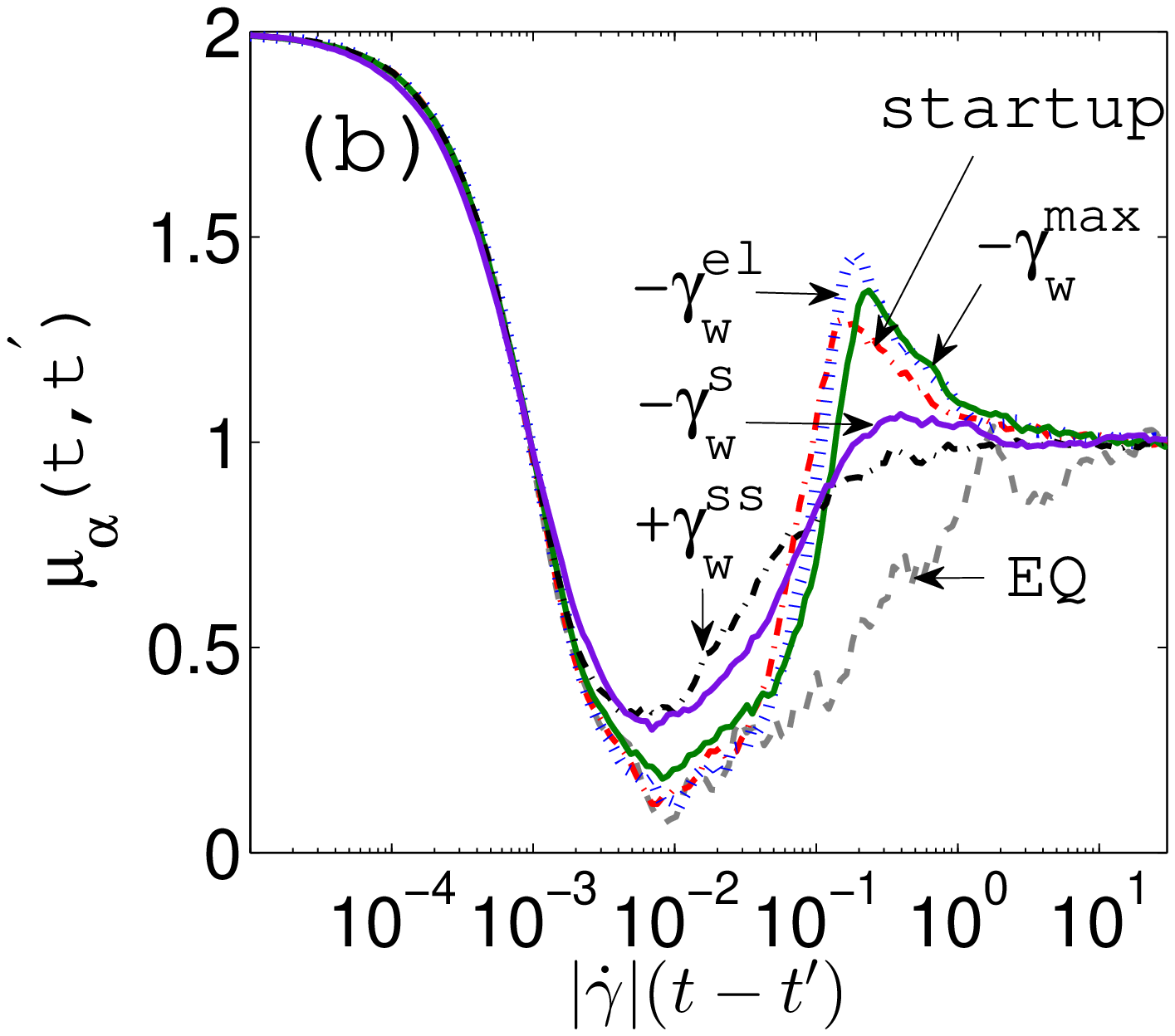}}
\caption{\label{fig:msd}
  (a) Mean-squared displacement of B particles in the vorticity direction.
  Shown are the equilibrium curve for $T=0.4$ (label EQ), and transient
  MSDs $\delta z^2(t,t')$ as a function of $|\dot\gamma|(t-t')$, where $t'$ is the
  time of initial flow start-up from equilibrium (EQ) respectively
  flow reversal according to the cases discussed in Fig.~\ref{fig:stressstrain}
  (steady-state, s; elastic transient, el; strain corresponding to the
  overshoot, max). The equilibrium curve is shifted to coincide at short times.
  (b) Logarithmic derivative $\mu_{t'}(t)=d\log\delta z^2(t,t')/d\log t$ for
  the cases shown in (a).}
\end{figure}

A related quantity that can be intuitively interpreted is the mean-squared displacement (MSD),
$\delta r^2(t,t')$, of a single particle. We choose randomly a particle of species B,
and show the results in Fig.~\ref{fig:msd}. Here we show, in addition to the cases
$t'\ll0$, $t'=0$, and $t'=t_w^\text{s}$ discussed above, also the MSDs obtained for
$t'=t_w^\text{el}$ and $t'=t_w^\text{max}$. For simplicity, we only record the
motion in the neutral direction; the MSD is then defined as $\delta r^2(t)=3\delta z^2(t)$
to match the spatially averaged MSD of the quiescent system at short times. $\delta z^2(t)$
is expected to become diffusive at long times, while the other components of the MSD under
shear attain superdiffusive asymptotes due to (the high-density analog of) Taylor dispersion
\cite{Krueger2011}.

It has been established by computer simulation and confocal microscopy studies \cite{Zausch2008,Zausch2010}
that the stress-overshoot in startup flow is accompanied
by superdiffusive particle motion on intermediate time scales.  
This is confirmed by the startup case shown in Fig.~\ref{fig:msd}: while the equilibrium
MSD remains subdiffusive at all times, the transient MSD obtained for $t'=0$ follows the
equilibrium one up to a time that corresponds to a few percent strain. Then, it quickly
crosses over to the diffusive regime obtained under steady-state flow.
An effective exponent can be assigned to the MSD curves by taking the logarithmic
derivative, $\mu(t)=d\log\delta r^2(t)/d\log t$. This quantity is shown in the
lower panel of Fig.~\ref{fig:msd}. Since the short-time motion of the simulation is
ballistic, $\mu(t)=2$ holds for $t\to0$; in the plateau region, $\mu(t)$ drops to
values close to zero, and for long times, $\mu(t)=1$ indicates diffusive motion.
For times where the fast increase in the $t'=0$ MSD is seen, $\mu(t)\approx1.3$
holds, in qualitative agreement with the simulation of Ref.~[\onlinecite{Zausch2008}].

Remarkably, the correlation between stress overshoot and superdiffusive motion of the
individual particles holds beyond the case of startup. In Fig.~\ref{fig:msd}, we observe
superdiffusive MSDs for those cases, where the corresponding stress--strain curve in
Fig.~\ref{fig:bauschinger} shows an overshoot; for the case of flow reversal in the
steady state, where the stress overshoot has vanished, also only subdiffusive motion
is seen in the MSD. This hints at a strong connection between the two phenomena. Such
a connection can be rationalized \cite{Zausch2008} by invoking a generalized Stokes-Einstein
relation: the mean-squared displacement is determined by a Mori-Zwanzig equation
similar to Eq.~\eqref{eq:eom}. The relevant memory kernel can be approximated by
the generalized time-dependent shear modulus appearing in the Green-Kubo relation for
the stress, Eq.~\eqref{eqn:greenkubo}. Although it does not provide a sufficient
condition, one sees that a memory kernel with negative portions in the MSD equations
may lead to superdiffusive behavior. Both stress-overshoot and superdiffusive MSD would then arise in the same time window characterized by the strain $\gamma^*$.

\section{\label{sec:4}Conclusions}

We have discussed the history-dependent response of sheared glass-forming
liquids subject to shear flow whose direction is instantaneously reversed.
Starting from the equilibrated liquid, application of a constant shear rate
$\dot\gamma$ causes the resulting stress $\sigma$ to increase from zero
to a steady-state value, undergoing an intermediate maximum, the stress
overshoot. Flow reversal, or equivalently, application of a constant shear
rate to a system that is in steady state corresponding to the flow in
opposite direction, results in a stress-versus-time curve that has no
maximum. At the same time, the (effective) elastic shear modulus found
from the initial rate of change of the stress is lowered, and the regime
in applied strain shrinks over which the behavior is described by this initial
linear elastic regime. These coincident phenomena can be seen as analogous
to the Bauschinger effect in simple shear flow. It should be pointed out that
the original Bauschinger effect \cite{Bauschinger} was discussed for metals
and steels under compressional and tensile load. However, given the generality
of history-dependent flow phenomena in amorphous systems, one may expect that
the underlying mechanisms are related to our discussion.

We have presented a systematic investigation of the gradual change between
the two cases of starting shear from a quiescent equilibrium and a
sheared steady state that belongs to the opposite flow. Reversing the
flow direction after an initial deformation that remains in the linear
elastic regime, no change in the stress overshoot or the elastic coefficients
is seen. The position of the overshoot indeed, as envisaged already by
Bauschinger \cite{Bauschinger}, marks the end of the reversible elastic
regime: at this point, plastic deformation gradually takes over, and
the effective elastic coefficients drop over a small window in
applied pre-strain.

Our theoretical analysis follows the integration-through-transients (ITT)
scheme upon which the mode-coupling theory for colloidal rheology is based.
The central physical mechanism expressed in our model is one of temporal
history dependence: glass-forming liquids are visco-elastic, with
a large time window $\tau$ over which the fate of past density fluctuations
affects the present response. It is this history dependence that causes
the Bauschinger effect and similar pre-shear-dependent phenomena in the model,
rather than spatial anisotropies induced by the pre-shear (since no
information on spatial variation of fluctuations is kept). This is in
alignment with the microscopic MCT and molecular-dynamics simulations of
the sheared thermal fluid \cite{Henrich2009,Krueger2011}, where it was found that the
steady state exhibits almost isotropic (one-time) pair correlation functions.
This may be different in the athermal limit.
It does, however, qualitatively explain the disappearance of the stress
overshoot, and our findings for the tagged-particle correlation functions
and mean-squared displacements. 
Immediately, this discussion predicts, that inserting a waiting time between the opposing flows  would cause the Bauschinger effect to weaken and go away. Any quiescent waiting period before shear-reversal would enable relaxation to proceed and would ultimately lead to startup curves from  the quiescent state.
When shear is started from the quiescent equilibrium
configuration, nearest-neighbor cages need to be broken before the accumulated strain
becomes effective in enhancing the relaxation dynamics. The critical strain is hence
connected to a typical cage size; some percent of the particle diameter following
a Lindemann-type criterion for melting. The sheared steady state is then characterized
by less strong cages, and the directionality of the flow does not play a major role
in this weakening effect. Hence, subsequent shear in the opposite direction will
lead to an earlier and more gradual relaxation of correlation functions
(cf.\ Fig.~\ref{fig:fsqt_shear}). The mean-squared displacements, observed to
follow the unsheared equilibrium curve up to times corresponding to the
critical strain in the case of startup shear, remain much closer to the
sheared steady-state ones in the case of flow reversal from such a steady state.
As a consequence, the sharp upturn observed for the startup case, connected to
superdiffusive motion and a stress overshoot, vanishes.

At present, the analysis of the initial stress overshoot
after startup of steady shear is based on a schematic model that does have
a number of fit parameters whose numerical ratio has to be determined
phenomenologically. However, the model is
rooted in an understanding of the microscopic MCT \cite{Amann2012}. In simplifying the latter
to arrive at a schematic model, one aims to keep those essential features
of the solutions qualitatively alike that the model shall describe; in this
sense, the model used here is a minimal model containing stress-overshoot
phenomena. It is important to note that the subsequent description of
the fate of the stress overshoot under time-varying flow drops out naturally
from the equations of motion. In this sense, the comparisons presented
in Figs.~\ref{fig:stressstrain} and \ref{fig:bauschinger} are qualitatively
different -- the first can be viewed as a model-motivated parametrization
of the simulation data. The second -- the comparison with the data on the
Bauschinger effect -- entails a parameter-free theoretical prediction.

\begin{acknowledgments}
We thank for funding by the Deutsche Forschungsgemeinschaft 
through Research Unit FOR 1394, projects P3 and P8.
A.~K.~B.\ acknowledges funding through the 
German Academic Exchange Service, DAAD-DLR programme.
Th.~V.\ is funded through the Helmholtz Gesellschaft (HGF, VH-NG~406)
and the Zukunftskolleg of the University of Konstanz.
\end{acknowledgments}
%

\end{document}